\documentstyle[12pt,epsfig,psfig]{article}
\begin{document}
\begin{center}

\baselineskip=24pt

{\Large \bf Energy calibration of large underwater detectors using
stopping muons}

\vspace{1cm}

{\large V. A. Kudryavtsev, R. A. Brook, S. L. Cartwright, J. E. McMillan,\\
N. J. C. Spooner, L. F. Thompson}

\vspace{0.5cm}
{\it Department of Physics and Astronomy, University of Sheffield, 
Sheffield S3 7RH, UK}

\vspace{1cm}
{\large \bf Abstract}
\end{center}

We propose to use stopping cosmic-ray muons in the energy calibration
of planned and deployed large underwater detectors. The method is based
on the proportionality between the incident muon energy and the length
of the muon path
before it stops. Simultaneous measurements of the muon path and 
the amplitude of
the signal from the photomultiplier tubes allow a relation
between the energy deposited in the sensitive volume of the detector and 
the observed signal to be derived, and also provide a test of detector 
simulations. We describe
the proposed method and present the results of simulations.
 
\vspace{0.5cm}
 
 
\vspace{1cm}
\noindent Corresponding author:  V. A. Kudryavtsev, Department of Physics
and Astronomy, University of Sheffield, Hicks Building, Hounsfield Rd., 
Sheffield S3 7RH, UK
 
\noindent Tel: +44 (0)114 2224531; Fax: +44 (0)114 2728079;
 
\noindent E-mail: v.kudryavtsev@sheffield.ac.uk

\pagebreak

\noindent {\large \bf 1. Introduction}
\vspace{0.2cm}

Several large neutrino telescopes have been proposed and are being deployed
deep under water or under ice \cite{Baikal,AMANDA,ANTARES,NESTOR}.
They are designed to measure the fluxes and spectra of high-energy 
astrophysical neutrinos by detecting muons produced via neutrino-nucleus
interactions. While these telescopes are designed to detect up-going
muons produced by neutrinos passing right through the Earth,
they can also detect down-going atmospheric muons. 
For most 
applications, such muons are a background which can be suppressed by
several orders of magnitude by the measurements of the arrival times
of Cherenkov photons and proper reconstruction of the muon track. These
muons, however, being of known origin with predictable flux and spectrum,
can serve also to test detector Monte Carlo and calibration
techniques. One possible application of atmospheric muons is the energy
calibration of the detector. 

The energy of a muon above 1 TeV can be derived from the number of collected
Cherenkov photons (converted to the amplitude or ADC channel). Since
for high-energy muons the mean energy loss is proportional to the muon path
and energy (stochastic energy losses dominate over ionisation loss), the
muon energy loss inside the detector volume (number of
produced Cherenkov photons) can be a measure of muon energy if the muon
path is reconstructed. 
This concept of muon energy measurement is proven 
by various simulations \cite{ANTARES,AMANDA1,Baikal1,ANTARES1}. 
The simulations are usually based on full
three-dimensional Monte Carlo simulations of the detector including 
muon energy loss,
production and propagation of Cherenkov photons, response of photomultiplier
tubes and electronics. The question is: how well can all the characteristics 
of all the elements of the detector be known or how accurately can they be 
measured with test light sources such as LEDs and lasers? 
Is it possible to check the simulation based predictions 
independently? In other words: can we know the muon energy from some other
independent source, which we can then compare with Monte Carlo predictions
and the measurement of the signal amplitude?

In the physics of keV-MeV energies there is a simple answer to a similar 
question. There are specific radioactive sources with mono-energetic 
X-ray or gamma-ray lines which are routinely used
for energy calibration of detectors. Only accelerators can provide a
quasi-monoenergetic particle beam at GeV-TeV energies, but
they are usually located far from large underwater or underground detectors.
Underground neutrino telescopes
and cosmic-ray muon detectors, being smaller in size (compared to underwater
detectors) and segmented structure, cannot measure the energy of
high-energy muons but can use their mean or peak energy losses together
with their path to calibrate the detectors (see, for example, \cite{VAK1}).
For the larger underwater detectors this method can hardly be applied
because the peak in the muon energy loss distribution is smoothed by
stochastic losses due to bremsstrahlung, pair production and inelastic
scattering over the long muon path.

We propose to use stopping atmospheric muons for this task. They have the
advantage that their incident energy can, in principle, be measured through
their path in the sensitive volume of the detector and, hence, can be
compared to the observed signal and predictions of simulations.
In Section 2 the basic idea of calibration is described and in Section 3 main
requirements for the detector hardware and software are given. Results of
our simulations for a simplified detector are presented in Section 4 and 
the conclusions are given in Section 5.

\vspace{0.5cm}
\noindent {\large \bf 2. Energy calibration of underwater detectors}
\vspace{0.2cm}

Usually energy calibration means the relation between particle energy
and the amplitude of the signal processed by DAQ. This definition is correct
if the particle energy is totally absorbed in the detector.
For underground detectors most muons are not absorbed in the detector
and the relation can be derived only between the observed signal and 
the energy deposited in the detector. 
Usually it is assumed that this relation is linear. 

Since there is no mono-energetic source of muons underwater, normal
methods of calibration cannot be applied. A mono-energetic source is not
necessary, however, if we can determine the muon energy from another measurement.
There is no way to measure the energy of individual muons from the whole
muon population, but there exists a sub-sample consisting of stopping muons 
for which
the incident energy can be known from the measurement of the muon path length
between the point where the muon enters the sensitive volume of the detector
and the point where it stops. The idea of the method is to measure
the muon path if it stops within the sensitive volume of the detector,
to calculate its initial energy using the energy-range relation and to compare
it to the measured signal amplitude, thus deriving the relation between the
amplitude and the deposited energy and checking Monte Carlo predictions.
Subsequently, the relation between the measured amplitude and the energy 
deposition can be
used to estimate the energy of through-going TeV muons.

One of the crucial points of the method is an estimation of incident
muon energy from the muon path length. Simple considerations show that, since
low energy muons (only such muons stop in the detector) mainly lose energy
via ionisation and atomic excitation without undergoing stochastic
energy loss, for most muons a simple linear relation will work
without large fluctuations. The results of the simulations will be shown in
Section 4.

Another important question is: can large underwater detectors
measure path length of a stopping muon with sufficient accuracy? This is 
discussed in the following section.

\vspace{0.4cm}
\noindent {\large \bf 3. Detector performance}
\vspace{0.2cm}

What are the performance requirements of an underwater detector for it
to be able to use stopping muons for energy calibration?

\noindent 1. Such a detector should be sensitive to down-going atmospheric 
muons. This is not necessarily true since neutrino telescopes have most 
(and possibly all) optical modules looking downwards.
Measurements of the depth-intensity relation for down-going muons
by the AMANDA and Baikal Collaborations \cite{AMANDA,Baikal2} prove that
down-going muons can be reconstructed and their fluxes can be measured.

\noindent 2. Separation between optical modules should not be large to allow
i) the detection of weak signals from low-energy muons and ii) the
determination of the stopping point of the muon with sufficient accuracy.
Requirement i) may reduce the sensitive volume of the detector
for this task but is unlikely to make the energy calibration impossible.
(Note, that the
Baikal energy threshold has been estimated as 10 GeV \cite{Baikal}
which is well matched to
our requirements). The requirement ii) is the most important. The distance
between optical modules on one string varies from 6 to 20 metres, while
the distance between strings is 20-80 metres for different arrays.
Simulations performed by the ANTARES Collaboration \cite{ANTARES,ANTARES1} for
neutrino oscillation studies show, however, that contained and semi-contained
neutrino-induced events can be successfully discriminated from through-going 
muons and, hence, both the 
points of muon production and absorption can be determined. Full 
reconstruction of muon events
is more difficult, of course, in the case of down-going muons because of
their large number. But even a fraction of stopping
muon events (well-reconstructed) may be enough for energy calibration.
(Note, however, that there is a danger of selecting a sub-sample of 
reconstructed
stopping muons with characteristics different from those of the whole
population of stopping muons).

\noindent 3. The rate of stopping muon events should be large enough to 
perform calibration in a feasible period of time. This requirement
may be more difficult to satisfy at deep sites. We will present our 
estimates in Section 4.

\noindent 4. Even if all requirements are satisfied, the proposed method 
cannot
avoid the use of detector simulation. This is due to the dependence 
of the observed
amplitude on the distance to the muon track. Imagine two vertical muon
tracks, one close to the string of optical modules and another 
far away from the string. Even if the lengths of the tracks are equal and
the incident muon energies are equal too, the observed signals will be
different due to different distances from the tracks to the phototubes.
Such simulations have already been performed by the collaborations
\cite{AMANDA,ANTARES,Baikal1}. The results depend strongly on the properties
of water/ice. The propagation of Cherenkov photons is not included in our
simulations, the assumption being that appropriate corrections for the 
distance between
optical modules and muon track can be made for a particular detector
if the track coordinates are fully reconstructed.

\noindent 5. Finally, multiple muon events can present a source of
background for stopping muons. Again, measured depth-intensity curves 
\cite{AMANDA,Baikal2} prove that at least some of the muon tracks can be
reconstructed properly.

As previously mentioned, there is a problem of possible bias in the
calibration if the sample of stopping muons that can be reconstructed has
particular characteristics different from those of the whole population
(for example, energy loss higher than the mean value). 
However, the probability of enhanced energy loss
along the whole muon track is extremely low. 
High amplitude photomultiplier hits due to occasional 
muon-induced local cascades can be excluded from the analysis. 
The presence of bias can be detected by comparing
the shape of the measured distribution of specific energy loss with 
the simulated distribution.

\vspace{0.4cm}
\noindent {\large \bf 4. Simulation of stopping muons}
\vspace{0.2cm}

In this section the results of simulations are presented which demonstrate 
that energy calibration, in principle, can be performed using stopping muons.
Further investigation of this question can be made for each
individual detector using appropriate software packages including
the detector design and characteristics.

The simulation was done in the following way: muons were sampled
according to their energy spectrum and
angular distribution at a particular depth underwater, propagated
through the detector volume and their initial and final characteristics
were stored on disk.
Energy spectra and angular distribution of muons under 
water were obtained by propagating muons with various initial energies
from sea level down to various depths in water using
the muon transport package MUSIC \cite{MUSIC}. Then, muon
energy distributions underwater were convolved with muon spectra
at sea level using the relevant parameterisation from \cite{Gaisser}.

To check the correctness of the procedure we calculated the muon
``depth -- vertical intensity'' relation and compared it with
those measured by the Baikal \cite{Baikal2} and AMANDA \cite{AMANDA}
experiments. The results are shown in Figure 1. Measured relations
are well fitted to the calculated relation assuming the power index of the 
primary spectrum is equal to 2.78 -- in good agreement with underground
experiments \cite{LVD,MACRO}. 

A special package was developed to sample single muons according to 
their spectra and angular distribution underwater: MUSUN (MUon Simulation
UNderwater/UNderground) \cite{VAK2}.
As an example, we present here the results for the depth of 2 km under water.
This corresponds approximately to the depth at which ANTARES will be deployed
\cite{ANTARES2} and
is 0.5 km deeper than the AMANDA site. A simple detector in the form of cylinder
with a radius of 100 metres and a height of 500 metres was used. This
cylinder represented the sensitive volume of the detector. Muons
were sampled uniformly on the upper surface of the cylinder. We required also
that the muon track (real or extended if the muon stopped) crossed the lower 
flat surface of the cylinder. It was assumed
that the muon track and stopping point can be accurately 
reconstructed if the muon traversed
between 100 metres and 400 metres of water inside the cylinder.

Figure 2 shows a scatter plot of initial energy versus track length for
stopping muons with track lengths between 100 and 400 metres. No
systematic uncertainty in track reconstruction was included.
There is a clear proportionality between track length and initial
energy with small fluctuations for most muons. Some of them, however,
undergo, a large stochastic energy loss.
Figure 2 can be converted into the distribution of specific energy
losses (per unit track length). This is shown by a dash-dotted line 
in Figure 3 for the same
stopping muon sample. The distribution is characterised by a narrow prominent
peak which corresponds to the most probable specific energy loss.
In practice, the track length can be reconstructed with some finite
accuracy. If we assume that the reconstructed track length distribution follows 
a Gaussian distribution with mean value equal to the true value of the track 
length and standard deviation equal to 40 metres
(at least twice the vertical distance between the modules on the string)
the peak becomes smoother but is still present (dashed line in Figure 3).
(Note, that the simulations of the ANTARES performance for neutrino
oscillation study \cite{ANTARES} show that the starting and stopping points
of more than half up-going muons can be reconstructed with an accuracy of
better than 30 m).

All previous results have been obtained assuming no fluctuations in the number
of detected photons due to various distances between the optical modules 
and muon track.
Such assumption is equivalent to the precise reconstruction of the position
of the muon track with respect to the optical modules 
and correction for such fluctuations.
It is unlikely, however, that this scenario is realised in practice. To
account for the fluctuations in the number of collected photons in our
simulations we sampled the reconstructed energy deposition according to
a Gaussian distribution on the log(E) scale with mean value equal to the
true energy deposition (which came as an output from the muon propagation code)
and standard deviation equal to 0.176 on the log(E) scale (factor 1.5). 
The choice of the standard deviation for simulations
was based on the known fact that the energy of
multi-TeV through-going muons can be reconstructed with an accuracy of 
0.4-0.5 on the log(E) scale \cite{ANTARES,Baikal1}. For high-energy
through-going muons, however, the energy resolution is determined mainly by
the energy loss fluctuations, while this effect (being much smaller
for low-energy stopping muons) is included separately in our simulations.
We included also the statistical fluctuations in the number of
detected photoelectrons assuming that on average an atmospheric muon
produces about 1 photoelectron per 1 GeV of energy deposition. 
To derive this number we used the mean number of
photoelectrons produced by an atmospheric muon in the Baikal
detector, quoted in \cite{Baikal} (23 photoelectrons),
the length of Baikal string which is equal to 72 metres, and the mean
energy deposition in water which is about 300 MeV/m.
We assumed also that the energy threshold of the detector is equal
to 20 GeV.
The results of the simulations are shown in Figure 3 by solid curve.
The distribution of specific energy losses is broader than in the
simpler case when we did not account for the fluctuations in
the number of detected photons. For comparison,
we present also a similar distribution for through-going muons which
pass more than 400 metres of water in the sensitive volume of the
detector. It is obvious, that the distribution of specific energy
losses for stopping muons is more favourable for the use in the
energy calibration procedure. This does not exclude, however, the
possibility to use such a distribution for the through-going muon
sample either as an additional test or in the case that our assumptions
about possible fluctuations were too optimistic.

The position of the peak calculated for each individual detector
can be used to convert the measured track length of a stopping muon to its
initial energy which can then be compared to the measured signal.
For our simple detector at 2 km depth the rate of stopping muons
which can be reconstructed is about 0.3 s$^{-1}$. 

Another important issue which we want to address in the discussion, is
the contamination from multiple muons. Multiple muons contribute from
a few percent to a few ten percent of the total muon flux even for
very large detectors. Their contribution decreases with depth.
Such events produce more light and hence, may be detected with
higher probability than single muons. If so, they can smooth
the observed distribution of specific energy losses.

To estimate the effect from multiple muons we simulated the
development of vertical Extensive Air Showers initiated by primary protons 
using the code CORSIKA
\cite{CORSIKA}. Then, we propagated all muons down to 2 km of water with
MUSIC and repeated simulations of muons (both single and multiple)
in the detector. The results are presented in Figure 4. Solid curve
shows the distribution of specific energy losses for single muons.
Note that the distributions shown by solid curves in Figures 3 and 4 are
very similar, though only vertical muons have been simulated with
CORSIKA. Then, we assumed that only events with total observed energy
more than 20 GeV, regardless of the muon multiplicity, can be detected.
We assumed also that multiple muon events could be reconstructed as 
stopping single muons only if all muons stopped inside the detector.
In this case 
the reconstructed track length corresponded to the muon with the longest
range. The results are shown in Figure 4 by dashed curve. From Figure 4
we conclude that multiple muons do not change significantly the
distribution of specific energy losses for stopping muons.
The contribution of multiple muons from heavy primaries and from
inclined directions can modify slightly calculated distribution
of specific energy losses but is is unlikely to change this conclusion.

It is possible that the number of Cherenkov photons
emitted by a low-energy muon is not enough to hit a phototube if the muon 
does not pass very close to it. In this case, only muons passing
very close to the strings can be used for calibration and a
single string analysis technique can be used (each string represents
a separate detector). The basic features of the method remain the same,
however, due to the reduced acceptance, the rate of reconstructed stopping
muons will be approximately 2 orders of magnitude less (if a 10 m radius
single string detector is used).

\vspace{0.4cm}
\noindent {\large \bf 5. Conclusions}
\vspace{0.2cm}

A method of energy calibration of large underwater detectors
using stopping muons has been described. 
The method is proven to work by Monte Carlo
simulations for a simple (``ideal'') detector. Further investigations
involving structure and characteristics of specific neutrino
telescopes are needed.

\vspace{0.4cm}
\noindent {\large \bf 6. Acknowledgements}
\vspace{0.2cm}

The authors wish to thank PPARC for financial support.
The idea of the method and first preliminary results of the simulations
were reported in May of 1998 on one of the meetings of the ANTARES
Collaboration. The authors are grateful to the Collaboration for
useful comments and discussion. One of the authors (VAK) 
thanks Dr. E. V. Korolkova for discussion and remarks.
We wish to thank also anonymous referee for useful comments.

\pagebreak

\begin{figure}[htb]
\begin{center}
\epsfig{figure=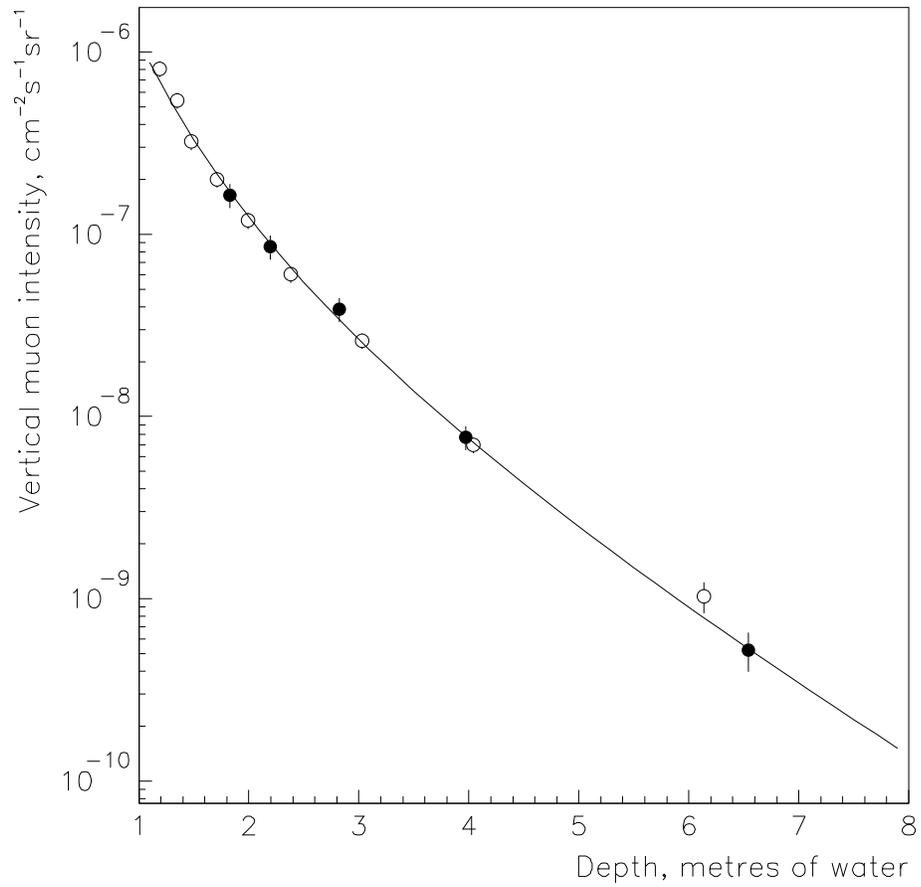,height=15cm}
\caption {Measured depth -- vertical muon intensity relation under
water fitted to the model described in the text. Data are from 
\cite{AMANDA} (filled circles) and \cite{Baikal2} (open circles).}
\end{center}
\end{figure}

\pagebreak

\begin{figure}[htb]
\begin{center}
\epsfig{figure=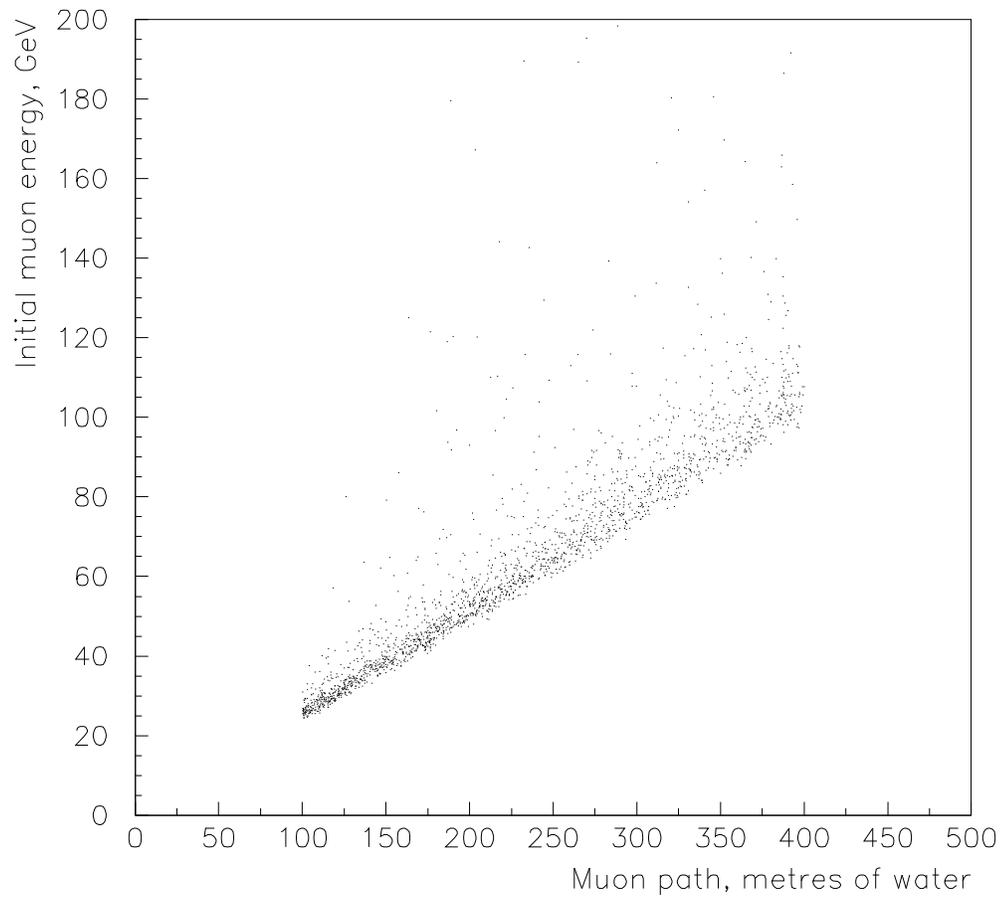,height=15cm}
\caption {Scatter plot of initial muon energy versus muon path for
muons stopping in the underwater detector (see text for details).}
\end{center}
\end{figure}

\pagebreak

\begin{figure}[htb]
\begin{center}
\epsfig{figure=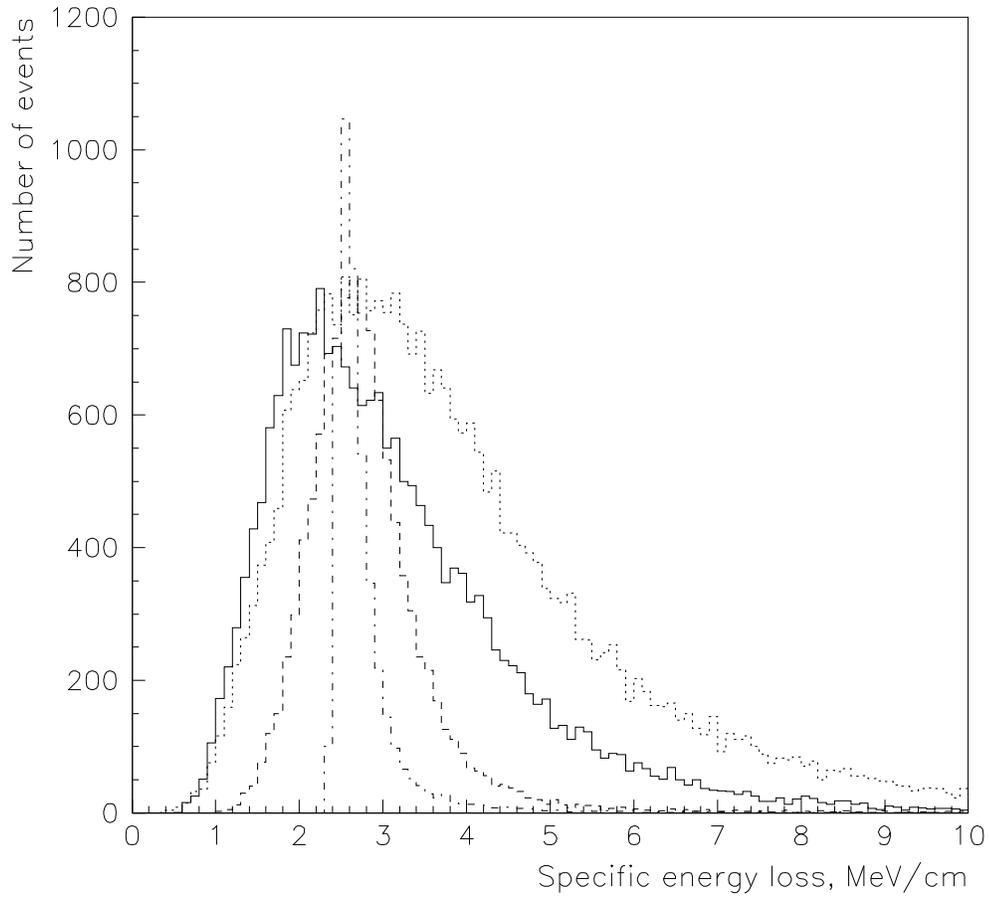,height=15cm}
\caption {Distribution of specific muon energy losses (energy losses
per unit track length). Dash-dotted curve -- stopping muons without uncertainty
in the track reconstruction and measured energy deposition
(scaled down by a factor of 5); dashed
curve -- stopping muons with 40 m uncertainty in the track length reconstruction
(scaled down by a factor of 2);
solid curve -- stopping muons with the uncertainties in the
measured energy deposition and track length;
dotted curve -- through-going muons with the uncertainties in the
measured energy deposition and track length
(see text).}
\end{center}
\end{figure}

\pagebreak

\begin{figure}[htb]
\begin{center}
\epsfig{figure=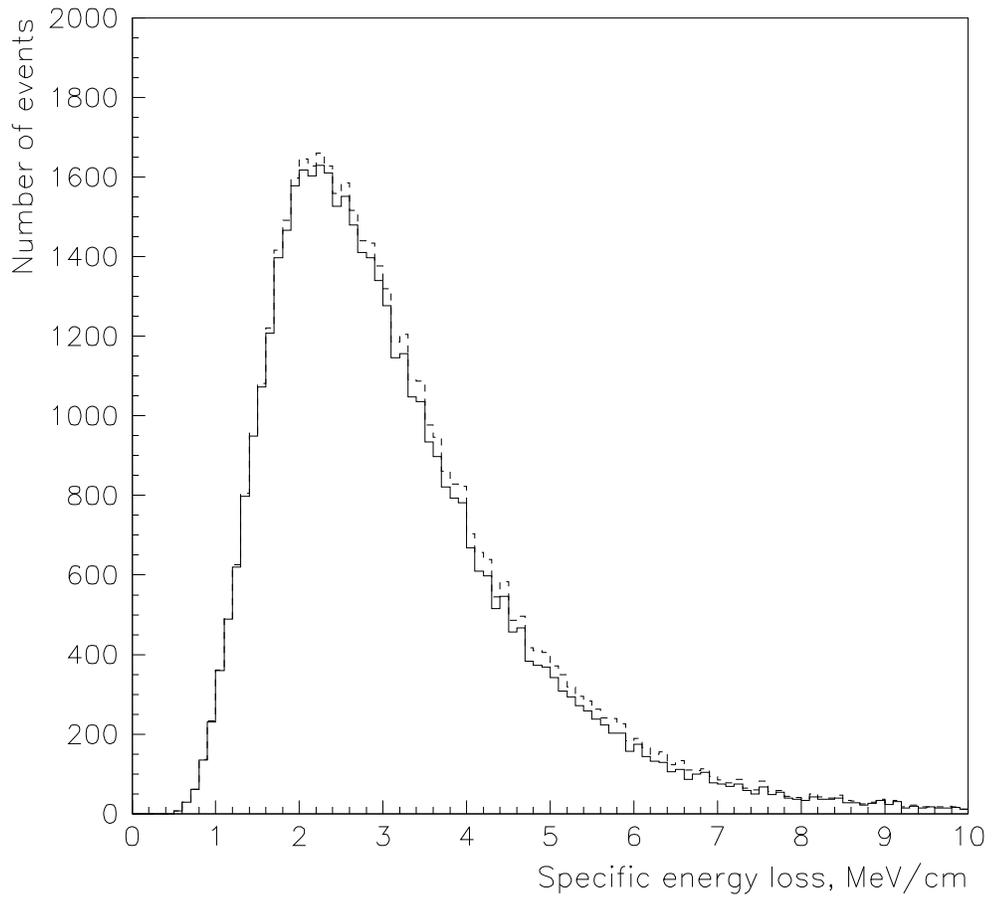,height=15cm}
\caption {Distribution of specific muon energy losses for
single muons (solid curve) and single+multiple muons (dashed curve).}
\end{center}
\end{figure}

\end{document}